\documentstyle[aps,prd,epsf]{revtex} 

\draft

\begin{document}
\twocolumn[\hsize\textwidth\columnwidth\hsize\csname
@twocolumnfalse\endcsname

\title{Galactic halos of self-interacting dark matter}

\author{Steen Hannestad}

\address{Institute of Physics and Astronomy,
University of Aarhus,
DK-8000 \AA rhus C, Denmark}

\date{\today}

\maketitle

\begin{abstract}
Recent, very accurate simulations of galaxy formation have revealed that
the standard cold dark matter model has great difficulty in explaining
the detailed structure of galaxies. One of the major problems
is that galactic halos are too centrally concentrated. 
Dark matter self-interactions have been proposed as a possible means
of resolving this inconsistency.
Here, we
investigate quantitatively the effect of dark matter self interactions
on formation of galactic halos. Our numerical framework is extremely
simple, while still keeping the essential physics. We confirm that
strongly self-interacting dark matter leads to less centrally
concentrated structures.
Interestingly, we find that for a
range of different interaction strengths, the dark matter halos
are unstable to particle ejection on a timescale comparable to the
Hubble time.
\end{abstract}

\pacs{PACS numbers: 95.35.+d, 98.62.Gq, 14.80.-j}
\vskip1.8pc]

\section{introduction}

The concept of dark matter was originally introduced by 
Zwicky in 1933 \cite{Z33}.
 to explain the behaviour of individual galaxy clusters.
Since then the dark matter has been determined to be essential
for explaining a vast number of astronomical observations.
The cold dark matter model has, with modifications, been very
successful in explaining how structure forms in the universe
\cite{gross}.
However, it was recently realised that the CDM model is apparently
unable to explain the dynamics of individual galaxies.
Very high resolution simulations have shown that there should
be much more sub-structure in the halos of typical galaxies than
is observed. 
Numerical simulations predict that the local group should at present
contain about 1000 distinct dark matter halos. Observations
yield a number which is a factor of ten lower
\cite{moore,ghigna}.
Also, lensing measurements of clusters seem to indicate a
constant density core, in strong contrast with numerical simulations
\cite{ghigna,tyson}. 
Also, the central density profile  in galactic halos is 
predicted to be much steeper than in real galaxies
\cite{FP94,N99,NS99,moore2,NFW96}.
However, this is not a completely settled issue. For instance 
Kravtsov {\it et al.} \cite{kravtsov} find that CDM simulations
are consistent with observations.

If the problem with CDM halos persists, then it
could possibly be remedied if structure formation
is somehow suppressed on small scales.
One such possibility is that perhaps the dark matter is not cold,
but rather ``warm'', i.e.\ dark matter particles have masses around
1 keV and therefore have significant thermal motion around the
time of matter-radiation decoupling \cite{SS88}. 
That would suppress structure formation on galactic scales and below.
However, warm dark matter
may have problems in describing properly the properties 
of clusters.
Another possibility is that the initial power spectrum has a
cut-off at some wavelength, corresponding to the substructures
in galaxies \cite{KL99}, so that no structure below
this scale will grow initially.

A quite different possibility is that the discrepancy has to do with
the interaction properties of the dark matter, not just the mass of the
individual particles. 
It was recently proposed by
Spergel and Steinhardt \cite{SS99}, that dark matter with
strong self-interactions could make the cold dark matter model
consistent with observations.
Strongly self-interacting dark matter will tend to produce
halos with 
shallower core density profiles, thereby alleviating the problem
of the central mass concentration.
Although the simple arguments provided in Ref.~\cite{SS99}
are quite convincing, it seems very important to test the 
implications of self interacting dark matter on a more
quantitative basis. That is the purpose of the present paper.
With the help of a very simple numerical scheme, we solve the
Boltzmann equation describing the phase-space evolution of
collisional dark matter and are thus able to calculate final
state density and velocity distributions. 

Intriguingly, we find that
there is a range of parameters for which there exist no stable
equilibria, even on a relatively 
short timescale. In general, however, we are
able to confirm the predictions of Ref.~\cite{SS99}, namely that
strongly self interacting dark matter produces halos with shallower 
density profiles and thus can remedy
 the problems that
the standard cold dark matter model has in explaining galactic
dynamics.
It should be noted here that self interacting dark matter has 
been considered previously, in order to study possible effects
on the initial linear power spectrum \cite{CMH92,LSS95,AD97}. 
However, the type of self 
interactions investigated in the present paper will have no
discernible effects on the initial power spectrum.


\section{The physics of self-interacting dark matter}

In the standard cold dark matter model, the dark matter particles
were once in thermal equilibrium in the early
universe. As the temperature dropped below their rest mass, however,
their abundance was exponentially suppressed. Eventually annihilation
reactions were no longer effective and the abundance of dark matter
particles froze out at some particular value \cite{KT90}. 
In this picture,
the dark matter particles are presently extremely weakly interacting
and halos consisting of them behave as collisionless systems.
However, dark matter particles could in principle have very strong
self-interaction, as long as their reaction cross section with
standard model particles is small. That is for instance the 
case with the non-thermally produced particles
 proposed by Chung, Kolb and Riotto \cite{CKR98}.
In the present paper we shall assume only that dark matter particles
are completely decoupled from all other species. The dark matter
self-interactions are modelled as point-like, i.e.\ the interactions
are mediated by some very massive boson. In that case, the cross section
for two-body scattering, $\sigma$, is energy-independent. This means
that the reaction rate for a dark matter particle moving in a dark
matter halo is
\begin{equation}
\Gamma = \int \frac{d^3 p}{(2\pi)^3} f_p v_{\rm rel} \sigma,
\label{eq:mfp}
\end{equation}
where $v_{\rm rel}$ is the relative velocity between the incident
particle and the target and $f_p$ is the phase-space distribution
of dark matter particles.
As will be explained later, the cross-section that divides the
very weakly interacting regime from the strongly interacting one
is given roughly by
\begin{equation}
\sigma_0 \simeq 1 \times 10^{-22} \, {\rm cm}^2 \, m_{\rm GeV}^{-1},
\label{eq:sigma}
\end{equation}
which of course is many orders of magnitude larger than the
interaction cross section for standard cold dark matter.


\section{Numerical scheme}

The fundamental equation describing the phase-space evolution
of self-interacting dark matter is the Boltzmann equation.
We shall in the present paper neglect the Hubble expansion
of the universe. In that case, the Boltzmann equation takes on
the form
\begin{equation}
\frac{df}{dt} = 
\frac{\partial f}{\partial t} + {\bf v} {\cdot} \frac{\partial
f}{\partial {\bf r}} + {\bf a} {\cdot} 
\frac{\partial f}{\partial {\bf v}} = C[f],
\end{equation}
where the right-hand side is a collision operator describing
possible scattering/annihilation reactions.

For normal, collisionless matter, the right hand side is zero, but
in our case it is a phase space integral of the reaction
matrix element.

In order to keep the numerical work quite simple, while still
retrieving the essential physics, we shall assume spherical 
symmetry of $f$. In that case, there are only four independent 
phase space coordinates, $(t,r,v_r,v_\perp)$. Instead of $v_\perp$ we
use the specific angular momentum, $j \equiv v_\perp r$, which 
is a conserved
quantity for collisionless systems \cite{fuji}. 
The left hand side
of the Boltzmann equation is then \cite{fuji}
\begin{equation}
\frac{df}{dt} = 
\frac{\partial f}{\partial t} + v_r \frac{\partial f}{\partial r}
+\left(\frac{j^2}{r^3} - \frac{G M(r)}{r^2}\right)
\frac{\partial f}{\partial v_r}.
\end{equation}

Our numerical scheme for solving the Boltzmann equation is somewhat
similar to that of Rasio, Shapiro and Teukolsky \cite{RST89}.
We mimic the phase space distribution by particles, distributed
according to the initial $f$. Then these particles are moved
in phase-space according to the Boltzmann equation.
Thus, the force on each phase-space ``particle'' is calculated
using the mean gravitational field produced by all particles,
$G M(r)/r$.
All mean field quantities, like $M(r)$, are calculated on an
Eulerian grid in phase-space and interpolated using spline
interpolation.
Thus, the code is a hybrid between a
real Boltzmann code and an N-body code. This is a standard way of
solving the collisionless Boltzmann equation, both in gravitational
dynamics and in plasma physics \cite{selwood,HE81}.

The right-hand side depends on the specific scattering process.
In the case of two-body scattering, it can be written as
\cite{KT90}
\begin{eqnarray}
C[f] & = & \frac{1}{2E}\int d^{3}\tilde{p}_{2}
d^{3}\tilde{p}_{3}d^{3}\tilde{p}_{4}
\Lambda(f,f_{2},f_{3},f_{4})
\label{integral}\\ 
& & \,\,\, \times \sum | \! M \! |^{2}_{12\rightarrow 34}\delta^{4}
({\it p}_{1}+{\it p}_{2}-{\it p}_{3}-{\it p}_{4})(2\pi)^{4}.
\nonumber
\end{eqnarray}
Here we have used the definition $d^{3}\tilde{p}\equiv
d^{3}p/((2 \pi)^{3} 2 E)$.
$\sum \mid\!\!M\!\!\mid^{2}$ is the interaction matrix element and
${\it p}_{i}$ is the four-momentum of particle $i$.
The phase-space factor is defined as $\Lambda \equiv
f_3 f_4 (1-f)(1-f_2)-f f_2 (1-f_3)(1-f_4)$.

However, since we are considering phase space to be sampled by
a distribution of particles, changing the phase space density due
to collisions amounts to letting the particles scatter on each other.

The interaction rate for a given particle
can therefore be written as in Eq.~(\ref{eq:mfp})
\begin{equation}
\Gamma = \int \frac{d^3 p}{(2\pi)^3} f_p v_{\rm rel} \sigma
\simeq n \sigma v_{\rm disp},
\end{equation}
where $v_{\rm disp}$ is the velocity dispersion of the system.
Therefore, in this simple approximation, the interaction rate
of a dark matter particle is given in terms of the simple mean
field quantities $n$ and $v_{\rm disp}$.

The above system of equations can be relatively easily solved for
given initial conditions, and quantities such as the radial
density distribution and velocity dispersion can be recovered.


\section{Results}

We have solved the system of equations described in the previous 
section for different values of the interaction cross section,
$\sigma$. The Boltzmann equation is non-dimensionalised in the
following way: Given a total mass $M_0$ of the system and a
characteristic radius $R_0$, the system is non-dimensionalised
by solving in terms of the dynamical time $\tau = \sqrt{R_0^3/GM_0}$
and the characteristic velocity $v_0 = \sqrt{GM_0/R_0}$. The 
interaction cross section should then be cast in units of
$\sigma_0 = m R_0^2/M_0$, where $m$ is the mass of the dark matter
particle. Note that for $\sigma = \sigma_0$, the mean free path
in a typical system should be $\lambda \simeq R_0$, so that that
$\sigma_0$ is a natural dividing line between slow and fast 
interactions.

In all the simulations we start with a homogeneous sphere
of radius $R_0$. The velocity dispersion is assumed to be
Maxwellian so that
\begin{equation}
f \propto \exp (-v_r^2/2 v_d^2-j^2/2 r^2 v_d^2).
\end{equation}
The velocity dispersion, $v_d$, is chosen to be $v_d = 0.1 v_0$.
If all particles start at rest, they will pass through the center
at the same time, and the simulation becomes unstable
\cite{bert}. However,
letting the particles have a small initial velocity does not alter
the final state of the whole system very much. It only
results in the collisionless systems being slightly less
centrally concentrated.

The system is allowed to evolve until $t=30 \tau$, in order to test
stability of the code. 

\subsection{Collisionless systems}

The first simulation was done for a normal collisionless dark matter
halo.
Fig.~1 shows how the mass distribution $M(r)$ evolves with time.
As expected, the system very quickly approaches equilibrium, essentially
in just one dynamical time.
The core profile of the relaxed system approximately follows a power 
law, $\rho \propto r^{-4/3}$.
Note that if the initial velocity of all particles was zero, the 
final system would have a density profile with $\rho \propto r^{-9/8}$
instead \cite{bert}. Thus, our assumption of a small initial
velocity leads to a less centrally concentrated halo structure.
However, since our purpose is to compare halos of dark matter
with different interaction strengths, and not so much to make exact
calculations, this assumption does not matter much.

\begin{figure}[h]
\begin{center}
\epsfysize=7truecm\epsfbox{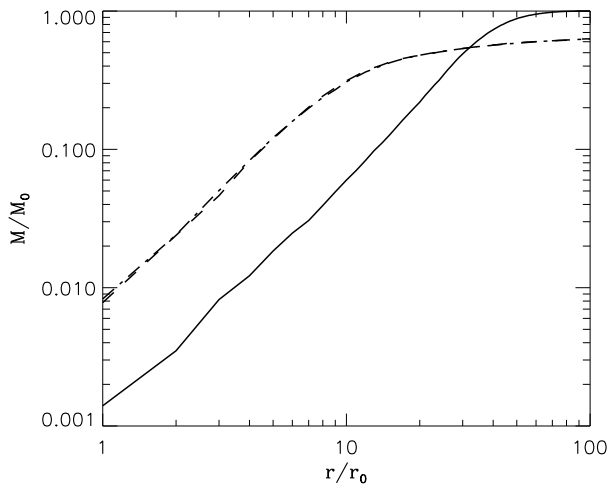}
\vspace{0truecm}
\end{center}
\baselineskip 17pt
\caption{The mass distribution, $M(r)$, for the collisionless dark matter
collapse at three different times.
The solid line is for $t=\tau$, the dashed for
$t=10\tau$, and the dot-dashed for $t=20 \tau$.}
\label{fig1}
\end{figure}

\subsection{Collisional systems}

{\it Ejection ---}
As described above, $\sigma = \sigma_0$ divides the weakly and strongly
interacting regimes. For $\sigma \ll \sigma_0$, one would expect an
evolution very similar to the fully collisionless system, whereas
for $\sigma \gg \sigma_0$, the system will behave like a collisional
gas, i.e.\ shocks can be produced and single particles will 
perform random walks in phase space.
The halo should therefore behave like an entirely
hydrodynamical core surrounded by a collisionless system.
However, at $\sigma \simeq \sigma_0$ the evolution is less obvious.
As also noted in Ref.~\cite{SS99}, most  self gravitating systems
can eject particles. In normal collisionless systems, this can
happen for two reasons, either a particle experiences a single
close encounter which leaves it with positive energy \cite{henon2}
or it suffers
many weak collisions, gradually increasing its energy \cite{henon1}. 
The first
process is usually referred to as ejection, whereas the second is
called evaporation. This process makes the entire system unstable
in the long term, since the only stable solution is a single 
pair of particles in a Kepler orbit, with all other particles
being at infinity \cite{BT87}. 
However, for a normal galaxy of stellar objects, the instability
time is vastly larger than the Hubble time, and for a collisionless
dark matter halo it is for all practical purposes infinite.

However, for a collisional halo, this need not be the case. Since 
particle interactions are point-like, a given particle can normally
only gain positive energy because of a single scattering event, not 
because of many weak collisions.
We can estimate the instability time in the following way:
In a homogeneous mass distribution with a typical velocity 
$v_0 = \sqrt{G M_0/R_0}$, a particle passing through the entire system
has the scattering probability
\begin{equation}
P \simeq 1-e^{-R_0/\lambda},
\end{equation}
where
\begin{equation}
\lambda \simeq \frac{m c}{\sigma \sqrt{\rho_0^3 R_0^2 G}}.
\end{equation}

If the particle scatters, and acquires a positive energy, then it should
not scatter again on its way out, if it is to be ejected. 
Therefore the total
probability for the particle to be ejected is roughly
\begin{equation}
P_{\rm ejection} \simeq (1-e^{-R_0/\lambda})e^{-R_0/\lambda}
P (E_{\rm final} > 0).
\end{equation}
$P (E_{\rm final} > 0)$ is the probability that the energy
after scattering is larger than zero. Normally this probability will
be $P (E_{\rm final} > 0) \simeq 0.1-0.2$, so that it is not 
negligible. The probability is maximal when $\lambda \simeq R_0$,
as could be expected. 
The typical timescale for instability is then roughly
\begin{equation}
t_{\rm instability} \simeq \tau/P_{\rm ejection},
\end{equation}
From these extremely crude estimates one finds that the minimal
possible instability time of of the order $t_{\rm instability}
\simeq {\rm few} \, \times \, \tau$.
Even if interactions are very strong, the system is unstable
in the long term, because particle will escape from the system
via diffusion \cite{SS99}. However, the timescale for this diffusion
process is much longer than the Hubble time.

\begin{figure}[h]
\begin{center}
\epsfysize=10truecm\epsfbox{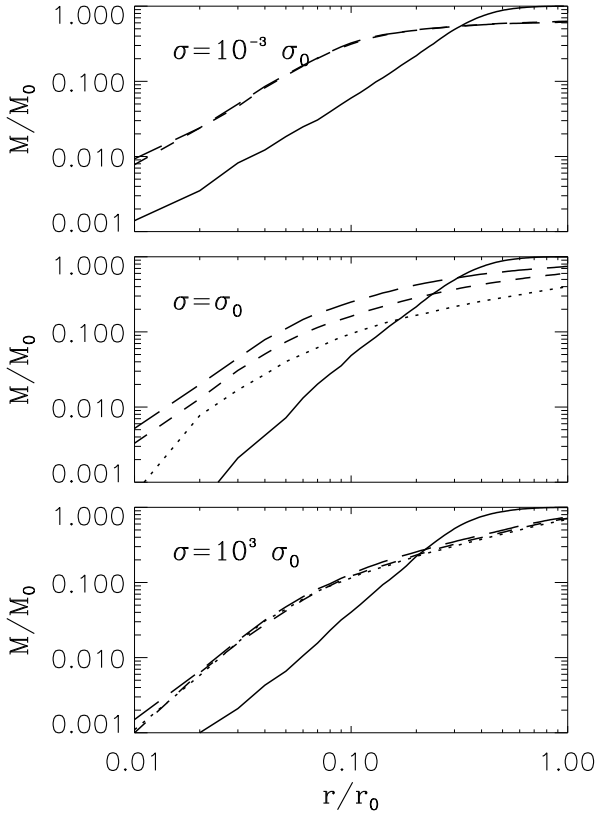}
\vspace{0truecm}
\end{center}
\vspace*{1cm}
\caption{The mass distribution as a function of $r$ at three different
times and three different values of $\sigma$. 
The solid lines are for $t=\tau$, the long-dashed for
$t=15\tau$, the dashed for $t=20 \tau$ and the dotted for $t=25 \tau$.}

\label{fig2}
\end{figure}

{\it Numerical simulations ---}
We have performed simulations for different values of $\sigma/\sigma_0$,
specifically $\sigma/\sigma_0 = 10^{-3},10^{-2},10^{-1},1,10,100,1000$.
Fig.~2 shows the evolution of $M(r)$ for $\sigma/\sigma_0 = 10^{-3}$,
1 and $10^3$.
For $\sigma/\sigma_0 = 10^{-3}$,
it is practically indistinguishable from the collisionless case,
as would be expected.
For the very strongly interacting case $\sigma/\sigma_0 = 1000$,
we find a very different behaviour. The system again settles
into equilibrium very fast, roughly at $t \simeq {\rm few} \times
\tau$, but
we also see that the core is much less dense because low entropy
material is ejected. This is the result predicted by Ref.~\cite{SS99}.

The intermediate case, $\sigma/\sigma_0 = 1$, never settles 
into a true long-term equilibrium. It starts out by approaching
the collisionless equilibrium, but scattering interaction drives
the mass distribution towards the equilibrium distribution for
the strongly interacting particles. However, on the same timescale
the system loses particles due to the ejection mechanism described
above, so that no true equilibrium is ever reached.
The timescale for particle loss indeed seems to be $\sim 10 \times 
\tau$.

{\it Observations ---}
We cannot expect our simulations to fit observational
rotation curves because of the approximations we have made,
i.e.\ Hubble expansion was neglected and spherical symmetry was
assumed. Nevertheless, it is interesting to compare our results
to observations. In Fig.~3 we show the predicted rotation curves at
$t = 15 \tau$ for the three different values, $\sigma = 10^{-3}$,
1 and $10^3$.
The rotation curve for the strongly interacting halo is much less
centrally peaked than the collisionless one.

The crosses in Fig.~3 show the measured rotation curve for the
typical low surface brightness galaxy UGC128 \cite{moore2}.
The observational rotation curve has an asymptotic circular 
velocity of $v = 200$ km/s at $R=100$ kpc. Taking $v_0=200$ km/s
and $R_0=100$ kpc gives a mass of $M_0 = 1.35 \times 10^{12} 
M_{\odot}$ and a dynamical timescale of $\tau = 4.7 \times 10^8$ y.
These numbers are typical for such galaxies. Notice that for this
choice of mass and radius, 
$\sigma_0 = 8.5 \times 10^{-23} {\rm cm}^2 m_{\rm GeV}^{-1}$,
i.e.\ corresponding to the $\sigma_0$ quoted in Eq.~(\ref{eq:sigma}).

\begin{figure}[h]
\begin{center}
\epsfysize=10truecm\epsfbox{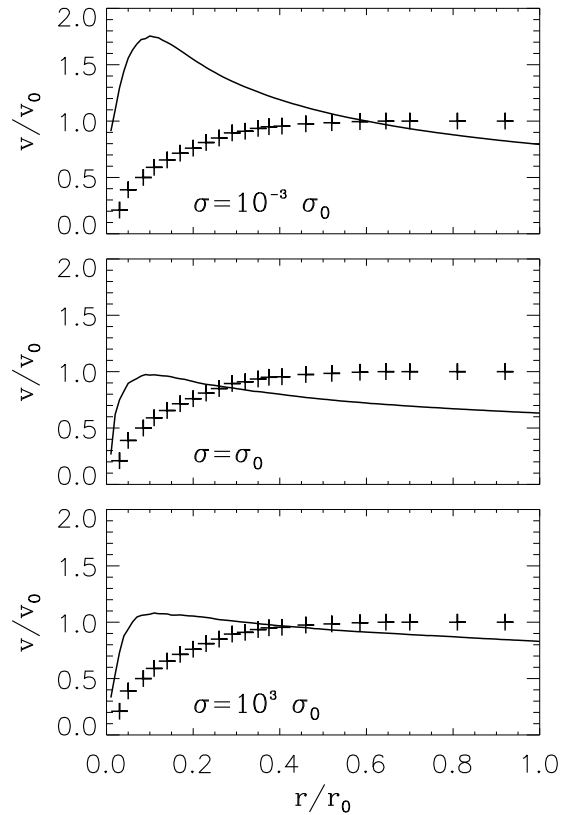}
\vspace{0truecm}
\end{center}
\vspace*{1cm}
\caption{The rotational velocity at $t=20 \tau$ for three different
values of $\sigma$. The crosses are the observational rotation curve
for the low surface brightness galaxy UGC128 \protect\cite{moore2}.}

\label{fig3}
\end{figure}

The rotation curve for the strongly interacting system provides
a better fit to the observational rotation curve than that
for the collisionless system. However, in all cases, the fits are
quite poor. As mentioned, this should not be taken too seriously 
since our model involves some essential approximations.


\section{Discussion}

We have performed quantitative calculations of how dark matter halos
form in models with self interacting dark matter. For simplicity
we assumed spherical symmetry, neglected the Hubble expansion,
and used a very simple prescription 
for the dark matter self-interaction. 

Our results have essentially confirmed the estimates of
Ref.~\cite{SS99}, in that dark matter halos with sufficient
self interaction will have much shallower density profiles that 
normal collisionless halos. Thus, they can provide better fits
to observational galactic rotation curves, while still being
consistent with all other known data.

A very curious feature is that halos with intermediate self interaction
($\lambda \simeq R_0$) are unstable on a rather short timescale
because they eject particles continuously. 
This effect was estimated to be unimportant in
Ref.~\cite{SS99}, but our numerical simulations indicate that there
exist regions of parameter space where the instability would probably
have shown up in present day halos. 

As mentioned, our calculations assume spherical symmetry
and neglect the Hubble expansion.
It will be
very interesting to investigate the effect of self interacting
dark matter using detailed N-body calculations \cite{KNH}.
Most likely, such simulations will confirm the general statements
made by Spergel and Steinhardt \cite{SS99}
as well as in the present paper.

\acknowledgements

Support from the Carlsberg foundation is gratefully acknowledged.


\end{document}